\documentclass[twocolumn,showpacs,preprintnumbers,amsmath,amssymb]{revtex4}

\usepackage{graphicx}
\usepackage{bm}
\usepackage{color}
\newcommand\Tm{\mathbf{T}}

\begin{document}

\title{Global mean first-passage times of random walks on complex networks}

\author{V. Tejedor$^{1,2}$}\author{O. B\'enichou$^{1}$}\author{R. Voituriez$^1$}

\affiliation{$^1$Laboratoire de Physique Th\'eorique de la Mati\`ere Condens\'ee
(UMR 7600), Universit\'e Pierre et Marie Curie, 4 Place Jussieu, 75255
Paris Cedex\\
$^2$Physics Department, Technical University of Munich, James Franck
Strasse, 85747 Garching, Germany}
\date{\today}

\begin{abstract}
We present a general framework, applicable to a broad class of random walks on complex networks, which provides a rigorous lower bound for the mean first-passage time of a random walker to a target site averaged over its starting position,  the so-called global mean first-passage time (GMFPT). This bound is simply expressed in terms of the equilibrium distribution at the target, and implies a minimal scaling of the GMFPT with the network size. We show that this minimal scaling, which can be arbitrarily slow for a proper choice of highly connected target,   is  realized 
under the simple condition that the random walk is transient at the target site, and independently of the small-world, scale free or fractal properties of the network.  Last, we put forward that the GMFPT to a specific target is not a representative property of the network, since the target averaged GMFPT satisfies much more restrictive bounds, which  forbid any sublinear scaling with the network size.
\end{abstract}

\maketitle

Complex networks have appealed a lot of  interest in the past few years \cite{Albert:2002,Dorogovtsev:2008,Barratbook}, mainly because of the extremely broad range of systems that they model, from biology to computer science or sociology. Despite this variety and their intrinsic topological complexity, many real networks have been shown to share some common features, such as the small-world property \cite{Watts:1998,Barratbook}, the scale free property \cite{Albert:1999a,Albert:2002}, or even fractal scalings \cite{Song:2005a}.  A crucial issue, still under  debate,  is to understand the impact of the topological complexity of such systems on transport properties. As a paradigm of dynamical processes, random walks on complex networks have been intensely studied \cite{0305-4470-36-1-304,Noh:2004a,Bollt:2005a,Samukhin:2008,Baronchelli:2008}, and in particular first-passage times have been widely used as a quantitative indicator of transport efficiency \cite{Huang:2006,Gallos:2007a}. The mean first-passage time (MFPT) to a target point was for instance calculated  in the case of fractal networks \cite{Condamin:2007zl,Benichou:2008a}.

Following the seminal work of Montroll \cite{Montroll:1969a}, many papers have  focused on the MFPT averaged over the starting point of the walker \cite{Kozak:2002,Bollt:2005a,Sood:2005,Agliari:2008,Haynes:2008,Zhang:2009b,Zhang:2009,zhang-2009a,agliari-2009}, sometimes called the  global mean first-passage time (GMFPT) .  Recently, a sublinear dependence on the size $N$ of the network of the GMFPT to  the most connected node of a specific network was  shown 
\cite{Zhang:2009},  and was interpreted as favorable for an efficient trapping. This finding,  in strong contrast with  previously known results in the case of regular  \cite{Montroll:1969a} or fractals \cite{Kozak:2002,Bollt:2005a,Agliari:2008,Haynes:2008} lattices, has motivated  an increasing number of works \cite{Zhang:2009b,Zhang:2009a,zhang-2009a,agliari-2009,zhang-2009e} that  have tried  to find examples of networks with high trapping efficiency, namely displaying weaker and weaker dependence on $N$ of the GMFPT.  Relying on these specific examples,  the heterogeneity, and more precisely the  scale free property  was put forward as advantageous  \cite{Zhang:2009b,Zhang:2009}, whereas the fractal property was suggested to be unfavorable \cite{zhang-2009a}.  

Here, we propose a general framework, applicable to a broad class of networks, which deciphers the dependence of the GMFPT on the network size $N$ and provides a global understanding of recent results obtained on specific examples \cite{Zhang:2009,Zhang:2009b,Zhang:2009a,zhang-2009a,agliari-2009,zhang-2009e}.
We first show on the example of  a new set of   networks that the GMFPT to the most connected node can scale as $N^{\theta}$, with $\theta$ {\it arbitrarily close to 0} despite the fractal property of the network.      
We then present an analytical approach which  yields (i) rigorous  bounds on the $N$ dependence of the GMFPT,  and (ii) a simple criterion under which this bound is reached, which in particular provides a condition for a sublinear scaling with $N$, which is independent of the scale-free, small-world, or fractal nature of the network.
Last, we show that a sublinear scaling is  never representative of the network, in the sense that the GMFPT averaged over the target site always scales faster than $N$.

\textit{Definition of the problem and notations.} We consider a set of graphs $\{{\cal G}_g\}_{g\in \mathbb{N}}$ where $N_g$ denotes the number of sites of the graph ${\cal G}_g$ at generation $g$,  such that $N_g\to\infty$ when $g\to\infty$.
We consider a discrete time random walker on ${\cal G}_g$. We assume that the transition probabilities $w_{ij}$ from site $i$ to site $j$ defining the walk are such that an equilibrium distribution $ P_{\textnormal{eq}}$ satisfying detailed balance exists. We further assume that  $\displaystyle\textnormal{sup}_{X\in {\cal G}_g}  P_{\textnormal{eq}}(X) \to 0$ when $g\to\infty$. We denote by $F_{S\to T}(\Tm=n)$ the probability that the walker reaches the target site $T$ starting from site $S$ for the first time after $\Tm=n$ steps, and write $ \overline{\Tm}_{S \to T} $ for the MFPT from $S$ to $T$. Note that this first average $\overline \cdot$ is taken over the realizations of the random walk. Taking the average of  the MFPT over the starting point, we define the GMFPT according to  : 
\begin{equation}\label{gmfpt}
\textnormal{GMFPT}(T)=\langle \overline{\Tm}_{S \to T}\rangle_{S \neq T}= \frac{\displaystyle \sum_{S \neq T}  P_{\textnormal{eq}}(S) \overline{\Tm}_{S \to T} }{1-P_{\textnormal{eq}}(T)}.
\end{equation}
 Note that  this quantity depends on the target point  $T$. Here the space average $\langle\cdot\rangle$ is taken over the equilibrium distribution $ P_{\textnormal{eq}}$, and slightly differs from the definition used in \cite{Zhang:2009c,Zhang:2009b,Zhang:2009,zhang-2009a,agliari-2009} where the average is taken over the flat distribution. It can be checked numerically on  networks recently studied in the literature that both definitions lead to the same scaling with $N_g$.

\begin{figure}[htbp]
\centerline{\includegraphics[scale=0.6]{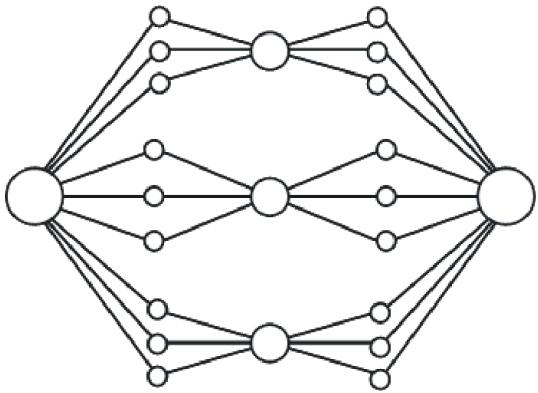}}
\caption{A fractal network leading to efficient trapping by the hub T (on the most connected sites): case of $k=3$, $u=2$ at generation $g=3$. 
}
\label{flowers2}
\end{figure}

 \textit{Efficient trapping on a fractal  network.} We first exhibit a set of fractal networks which extends the so-called $(u,u)$-flowers introduced in
\cite{Rozenfeld:2007}, and whose 
GMFPT to the most connected node  scales as $N^{\theta}$, with $\theta$ arbitrarily close to 0.  The first generation of the graph consists in two nodes connected by one
link ; then, at each
iteration, every link is broken and replaced by $k$ paths of $u\ge 2$ links
({\it cf.} figure \ref{flowers2}). It is clear that this network is fractal with a fractal dimension $d_f=\ln(ku)/\ln(u)$, since the  diameter  of the network at generation $g$ is  $L_g \sim
u^g$ while the number of sites is $N_g
\sim (ku)^g$ (the usual $(u,u)$ flowers of \cite{Rozenfeld:2007} correspond to the special case $k=2$). Taking the  target as one of the initial node, it is easily seen that the determination of the
 GMFPT is  actually a 1D problem since  all the points $n(r)$ at the same distance $r$ of the target are equivalent by symmetry, and thus lead all to the same $\overline{\Tm}(r)$.
Noting next that for 
all $r \in [1,u^g-1], \: P_{\textnormal{eq}}(r) n(r) = 2
P_{\textnormal{eq}}(T)$ and $P_{\textnormal{eq}}(u^g) n(u^g) =
P_{\textnormal{eq}}(T)$, and using  the  classical 1D expression 
$\overline{\Tm}(r) = r (2 u^g-r)$ \cite{Redner:2001a}, we obtain the following exact expression: $\textnormal{GMFPT}(T) =  \frac{2u^{g} ( 2u^{g}+1)}{6}\propto N^{\ln(u^2)/\ln(ku)}$.
In other words, for $k$ large enough,  the trapping  at the hub  is {\it arbitrarily efficient}  on this  network despite its fractal property. 

\textit{Lower bound of the GMFPT.} In order to gain  understanding in the real parameters relevant  to the scaling of the GMFPT with the size $N$, we now derive a general  lower bound for the GMFPT.  This derivation follows from the generalization of the Kac formula \cite{Aldous:1999,Condamin:2005db}  which we briefly recall here for the sake of selfconsistency.
We start from the following backward equation satisfied by $\displaystyle{F_{S\to T}}$ for $n\ge 2$ (see \cite{Redner:2001a}):   
$F_{S\to T}(n)=\, \sum_{j\neq T}\, w_{Sj} \, F_{j\to T}(n-1)$,   
which is completed by  $F_{S\to T}(n=1)=w_{ST}$. Laplace transforming and averaging this equation over $S$ (with a weight $ P_{\textnormal{eq}}(S)$ as in Eq.(\ref{gmfpt})) 
yields the generalized Kac formula
\begin{equation}
 \label{Kac}
\frac{ P_{\textnormal{eq}}(T)}{1-P_{\textnormal{eq}}(T)}\left( \hat{F}_{T\to T}(s)-e^{-s}\right)=(e^{-s}-1) \langle \hat{F}_{S\to T}(s)\rangle_{S\neq T},
\end{equation}
where  $ \displaystyle\hat{F}_{S\to T}(s)\equiv\sum_{n=1}^\infty e^{-sn} F_{S\to T}(n) $.
This very general equation, derived in a similar form in \cite{Condamin:2005db},  relates the distribution of the first return time to a site $T$ to the distribution of the global first-passage time to $T$. Expanding Eq.(\ref{Kac}) to  first order in $s$  yields the classical Kac formula $\overline{\Tm}_{T\to T}=1/ P_{\textnormal{eq}}(T)$ \cite{Aldous:1999,Condamin:2005db}. In turn, the second order in $s$ gives :
\begin{equation}\label{GMFPTexact}
\textnormal{GMFPT}(T) = \frac{1}{2} \frac{P_{\textnormal{eq}}(T)\overline{\Tm^2}_{T \to T}  - 1}{1 - P_{\textnormal{eq}}(T)}.
\end{equation}
Using next $\overline{\Tm^2}_{T \to T}  \ge  \overline{\mathbf{T}}_{T \to T} ^2 $ and the classical Kac formula,  the above exact expression gives a lower bound for the GMFPT:
\begin{equation}
\textnormal{GMFPT}(T)  \geq \frac{1}{2 P_{\textnormal{eq}}(T)}.
\label{GMFPTineq}
\end{equation}
Note that this lower bound is in close analogy with the one obtained in \cite{Benichou:2005a} in the context of continuous space Pearson random walks in confinement.

We now discuss under which conditions this lower bound is reached.
Strictly speaking, this requires the very restrictive condition that the variance of $\Tm_{T \to T}$ is zero. More generally we can discuss under which conditions the right and the left hand side of Eq.(\ref{GMFPTineq}) share the same scaling in the large size limit. To do so, we consider a sequence of target sites $\{T_g\in {\cal G}_g\}_{g\in \mathbb{N}}$, which can be for instance hubs of the networks at each generation as in refs  \cite{Zhang:2009b,Zhang:2009,agliari-2009}. Using (\ref{GMFPTineq}), and recalling that we have assumed $P_{\textnormal{eq}}(T_g) \to 0$ for $g\to \infty$, we define the minimal scaling of the GMFPT for $g\to \infty$ by
\begin{equation}\label{minscaling}
\textnormal{GMFPT}(T_g)=O(1/P_{\textnormal{eq}}(T_g)).
\end{equation}
Eq. (\ref{GMFPTexact}) then shows  straightforwardly that this minimal scaling  
is realized as soon as the  reduced variance of the first return time is finite in the large size limit, namely:
 $(\overline{\Tm^2}_{T \to T}    -  \overline{\mathbf{T}}_{T \to T} ^2 ) /\overline{\mathbf{T}}_{T \to T} ^2 = O(1)$.

We now show that this condition  for a  minimal scaling with the network size $N_g$  is actually equivalent to the transience property of  the random walk at the target site $T_g$ in the large size limit.
We first derive an alternative exact expression for the GMFPT. Let us introduce the pseudo-Green functions \cite{Barton:1989a,Condamin:2005db,Condamin:2007zl}  defined as:
\begin{equation}\label{H}
H_{S\to T} = \sum_{n=1}^{\infty} \!\left( P_{S\to T}(n) - P_{\textnormal{eq}}(T) \right),
\end{equation}
where $ P_{S\to T}(n)$ is the propagator, namely the probability that the walker is at  $T$ at time $n$ starting from $S$. 
It can be shown (see \cite{Noh:2004a,Condamin:2005db,Condamin:2007zl})  that the MFPT is then given by the exact expression:
$\overline{T}_{S \to T}  =  \frac{1}{P_{\textnormal{eq}}(T)}\left ( H_{T\to T} -  H_{S\to T} \right )$.
Making  use of the  relation  $P_{\textnormal{eq}}(S) H_{S\to T} = P_{\textnormal{eq}}(T) H_{T\to S}$, which follows from detailed balance (see also \cite{Noh:2004a}), we  obtain a second exact expression for the GMFPT:
\begin{equation}\label{exact2}
\textnormal{GMFPT}(T) = \frac{ H_{T\to T}}{P_{\textnormal{eq}}(T) \left ( 1- P_{\textnormal{eq}}(T) \right )}.
\end{equation}

This equation provides an alternative condition under which  the minimal scaling
 is realized, given by  $H_{T\to T} = O(1)$ in the large $g$ limit. From the definition (\ref{H}) of $H_{T\to T}$, this condition states that  the random walk is transient at site $T_g$  in the limit $g\to \infty$, i--e that in this limit, a random walker returns on average only a \textit{finite} number of times to $T_g$ \cite{Burioni:2005}. Conversly, Eq. (\ref{exact2}) indicates that if the walk is recurrent at $T_g$ for $g\to \infty$, that is if $H_{T_g\to T_g}$ diverges for $g\to \infty$, then  $\textnormal{GMFPT}(T_g)$ grows faster than $1/P_{\textnormal{eq}}(T_g)$. 

The lower bound (\ref{GMFPTineq}) and  minimal scaling (\ref{minscaling}) for the GMFPT  obtained above  call for  comments. (i) First, our analysis puts forward a very general criterion to have a minimal scaling of the GMFPT with the network size, namely the  type (transient or recurrent) \cite{Burioni:2005} of the random walk at the target site. We stress that this criterion is independent of the scale-free, small world or fractal properties of the network. Note that for a generic set of graphs $\{{\cal G}_g\}_{g\in \mathbb{N}}$, the type of the random walk for $g\to \infty$ is a site dependent property  \cite{Burioni:2005,Bollt:2005a,Rozenfeld:2007}. (ii) Second, the minimal scaling (\ref{minscaling}) is fully determined by the equilibrium distribution at the target site, which is generally much easier to obtain than dynamical quantities, and which crucially depends on the connectivity of the target site. Let us take the classical example of an isotropic random walk, for which $w_{ij}=1/k_i$ if $i$ and $j$ are neighbors and else 0, where $k_i$ denotes the connectivity of site $i$.  The minimal scaling of the GMFPT to a target $T_g$  then reads  $N_g\langle k\rangle/k_{T_g}$,  where $\langle k\rangle$ is the connectivity averaged over all sites. 
(iii) Note finally that in the case of a recurrent random walk at the target the minimal scaling is not realized, but the scaling of the GMFPT can however be sublinear if the growth of the connectivity at the target is fast enough. In this case  the scaling of the GMFPT depends on the scaling of $H_{T_g\to T_g}$, which generally depends both on the network
and on the target  $T_g$.

It is noteworthy that  our analysis provides a comprehensive view of recent papers highlighting a sublinear dependence of the GMFPT to a hub on different examples of networks. (i) In the example of deterministic scale-free graph proposed in \cite{agliari-2009},  the minimal scaling that we predict in Eq.(\ref{minscaling}) is indeed realized and the transience of the random walk at the target site (as defined above) is shown in the limit of large size 
(since the probability to come back at the hub in a finite time is null, as can be seen from Eq.(36) from \cite{agliari-2009} in the large size limit), in agreement with our approach. (ii) The authors of \cite{Zhang:2009b,Zhang:2009} have studied different examples of small world  scale-free networks (Apollonian networks \cite{Zhang:2009b} and $(u,v)$ flowers \cite{Zhang:2009}) where the GMFPT to the main hub displays a sublinear scaling. In these examples the scaling of  $\textnormal{GMFPT}(T_g)$ is strictly faster than our predicted minimal scaling  $1/P_{\textnormal{eq}}(T_g)$ (and satisfies the upper bound given in (\ref{2bounds})). Our criterion therefore implies that random walks on such structures are recurrent at the target site in the large size limit. 

\textit{Bounds on the average GMFPT.} As demonstrated previously, the GMFPT highly depends on the  target site, especially in the case of  scale-free network where the connectivity can be very heterogeneous. Therefore the   GMFPT to a specific target site cannot be taken as a general characteristic of the network. Actually, as we proceed to show the GMFPT averaged over the target site, defined by
$\langle \textnormal{GMFPT} \rangle = \sum_{T} P_{\textnormal{eq}}(T) \textnormal{GMFPT}(T)$,
has scaling properties with $N_g$ which can widely differ from the case of a fixed target site studied above. 
The inequality (\ref{GMFPTineq}) gives straightforwardly the following lower bound for $\langle \textnormal{GMFPT} \rangle$ (see aso \cite{Aldous:1999}):
$\langle \textnormal{GMFPT} \rangle \geq \frac{N_g}{2}$.
Hence, the averaged GMFPT always scales faster than $N_g$, and sublinear scalings discussed above are pointwise properties which are {\it never} representative of the network. This general inequality 
sheds some light on the result obtained by Bollt and ben Avraham \cite{Bollt:2005a} in the case of a specific network ((1,2) flowers), where the GMFPT averaged over a fraction of nodes of the network scales sublinearly with $N_g$, while
the GMFPT averaged over all the nodes is linear. 
Interestingly,  we can also propose an upper bound for $\langle \textnormal{GMFPT} \rangle$ following \cite{Aldous:1999}. First we define
 (see also \cite{Bollt:2005a,Baronchelli:2008}) 
the mean commute time as:
$\tau_{ij}=   \overline{ T}_{i \to j}  + \overline{ T}_{j \to i} $.
The quantity $\tau_{ij}$ can actually be bounded using a very useful  electrical analogy. Let us assign a unitary resistance to each link of the graph. Then it can be shown (see \cite{Chandra}) that the following general relation holds
$\tau_{ij} = N_g \langle k \rangle  r_{ij}$,
where $r_{ij}$ is the effective electrical resistance of the network between sites $i$ and $j$. It is then straightforward to obtain that $r_{ij}\le d_{ij} $ where  $d_{ij}$ is the distance between $i$ and $j$.  Indeed, $d_{ij}$ is the resistance of a path of length $d_{ij}$ between $i$ and $j$, and any parallel paths can only lower the resistance. We therefore finally obtain:
\begin{equation}\label{bounds}
\frac{N_g}{2} \leq \langle \textnormal{GMFPT} \rangle \leq \frac{N_g \langle k \rangle \langle d\rangle}{2},
\end{equation}
where $\langle d\rangle$ is the weighted average over pairs of the point to point distance $d_{ij}$.

Importantly, this shows that the scaling of  $\langle \textnormal{GMFPT} \rangle$ is much more constrained than the scaling of the GMFPT for a fixed target. This is particularly striking in the case of small world networks for which $\langle d\rangle\sim  \ln N_g$  : hence in case of  small-world networks with finite $\langle k \rangle$, widespread in nature \cite{Barratbook}, this shows that   $\langle \textnormal{GMFPT} \rangle$ always scales linearly with $N_g$ (up to log corrections). Note also that these bounds (\ref{bounds}) are compatible with
the linear scaling of  $\langle \textnormal{GMFPT} \rangle$ with $N_g$ reported in the case of Apollonian networks \cite{Huang:2006} and (1,2) flowers  \cite{Bollt:2005a}. The conditions for which the scaling of each of the bounds in (\ref{bounds}) is realized can also be discussed. As for the scaling of the lower bound, a sufficient condition for its realization is that for any sequence of targets $\{T_g\in {\cal G}_g\}_{g\in \mathbb{N}}$, the random walk is transient at $T_g$ in the limit $g \to \infty$. Note however that this condition is not necessary, and the bound can be reached for networks having mixed type properties, as in the case of (1,2) flowers already mentioned \cite{Bollt:2005a}. As for 
the scaling of the upper bound,  first notice that for any tree graph,  $r_{ij} $ is exactly  the distance $d_{ij}$ as discussed above using the electrical analogy. We conclude that for any tree the scaling of the upper bound is realized.  In particular we find that $\langle \textnormal{GMFPT} \rangle\sim N_g\ln N_g$ for {\it any} small world tree (see \cite{zhang-2009d} for an example).

Additional comments are in order. (i) First,  Eq.(\ref{bounds}) provides as a by-product an upper bound for the GMFPT itself, leading finally to:
\begin{equation}\label{2bounds}
\frac{1}{2 P_{\rm eq}(T)}\leq \textnormal{GMFPT}(T)  \leq  \frac{N_g \langle k \rangle \langle d\rangle}{2 P_{\rm eq}(T)}.
\end{equation}
(ii) Second, this upper bound for $\textnormal{GMFPT}$  inductively gives an upper bound of the trapping time in the case of a moving target using the Pascal principle \cite{Moreau:2003xe}. (iii) Last, we underline that in the case of fractal networks, characterized by a fractal dimension $d_f$ and a walk dimension $d_w$ \cite{Song:2005a} an explicit scaling of $\langle \textnormal{GMFPT} \rangle$ can be obtained  (see  \cite{Bollt:2005a,Agliari:2008}). Indeed, 
using for instance the asymptotics  of  the MFPT between points separated by a distance $r$  \cite{Condamin:2007zl} and averaging over $r$, one gets the following scaling 
\begin{equation}
\langle \mbox{GMFPT} \rangle \sim  \left \{
\begin{array}{ll}
N_g & \mbox{if } d_w < d_f\\
N_g \ln (N_g) & \mbox{if } d_w = d_f\\
N_g^{d_w/d_f} & \mbox{if } d_w > d_f\\
\end{array}
\right .,
\label{OetR}
\end{equation}
 which depends on the type of the random walk (transient if $d_f>d_w$).

\textit{Conclusion. }
We have presented  a general framework, applicable to a broad class of networks, which provides  rigorous  bounds on the size dependence of the GMFPT to a target site.  We have shown that the GMFPT has the same scaling in the large size limit as this lower bound under the condition that the random walk is transient at the target site. This shows that the type of the random walk (transient or recurrent) is a crucial criterion to determine the scaling of the GMFPT, widely independent of its scale free, small world, or fractal properties.  This study reconciles recent works on GMFPT for random walks on various network examples. Additionaly, we have demonstrated that the scaling of the GMFPT to a specific target is not a representative property of the network, since the target averaged GMFPT satisfies much more restrictive bounds, which in particular forbid any sublinear scaling with the network size.

We thank E. Agliari and R. Burioni for useful discussions.



\end{document}